\documentclass[11pt]{article}

\usepackage[english]{babel}
\usepackage[utf8]{inputenc}
\usepackage[T1]{fontenc}
\usepackage[a4paper, margin=3.5cm]{geometry}
\usepackage{amsmath, amsfonts, amssymb, amsthm, subcaption, hyperref}
\usepackage{graphicx}
\usepackage{authblk}
\usepackage{lmodern}

\newtheorem{definition}{Definition}

\usepackage[backend=biber, style=alphabetic, maxbibnames = 5]{biblatex} 
\graphicspath{{./figures/}}

\newcommand{\Nant}{\boldsymbol{\uparrow}}
\newcommand{\Eant}{\boldsymbol{\rightarrow}}
\newcommand{\Want}{\boldsymbol{\leftarrow}}
\newcommand{\Sant}{\boldsymbol{\downarrow}}

\newcommand{\turnleft}{\mathrm{turn^\circlearrowleft}}
\newcommand{\turnright}{\mathrm{turn^\circlearrowright}}

\begin{document}
\title{The $LLLR$ generalised Langton's ant}
\author[1*]{Victor Lutfalla}
\affil[1]{Université Publique, France}
\affil[*]{\tt victor.lutfalla@math.cnrs.fr}
\date{2025}
\maketitle

\begin{abstract}
  We present a short note on the dynamics of the $LLLR$ generalised Langton's ant.
  We describe two different asymptotic behaviours of the $LLLR$ ant.
\end{abstract}

\paragraph{Keywords.} Langton's ant, emergent behaviour, highway conjecture, turmites

\paragraph{Acknowledgements.} I thank Michael Rao who presented to me the facts I formalise here and who performed the intensive computation I mention at the end of this note. 

\section{Introduction}
The original \emph{Langton's ant} model was introduced in the 80's by different researchers \cite{langton1986, dewdney1989} as a model for ``artificial life'' meaning a simple discrete dynamical system exhibiting some kind of \emph{emergent behaviour}.

\begin{figure}[h!]
  \center
  \includegraphics[width=0.5\textwidth]{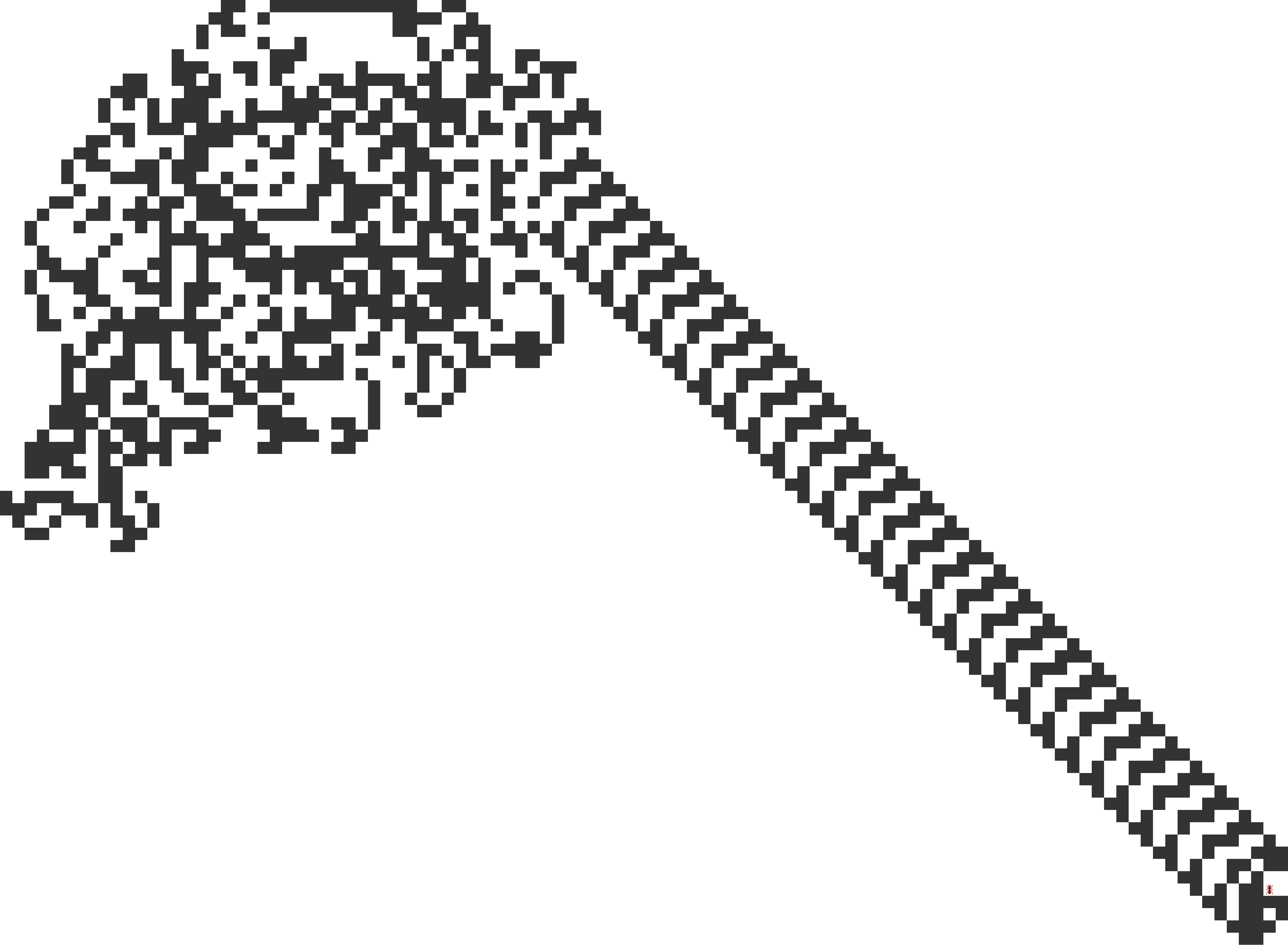}
  \caption{The configuration reached by Langton's ant from the uniform configuration after $13\,000$ steps. We can observe a periodic pattern leaving the initial seemingly chaotic pattern.}
  \label{fig:LR_highway}
\end{figure}

This emergent behaviour, called the \emph{highway}, is periodic with a drift.
From the initially uniform configuration, Langton's ant has, at first, a seemingly chaotic behaviour and after almost $10\, 000$ steps it enters the highway behaviour described above, see Fig.~\ref{fig:LR_highway}.

It was conjectured that the same asymptotic behaviour, the highway of period $104$ and drift $(\pm 2, \pm 2)$, is reached from any initial configuration with a finite number of non-zero cells.

Several generalisations have been proposed for this model.
The generalisation that seems to better preserve the original ant's properties is the one called \emph{generalised ants}~\cite{gale1994,gale1998}.
In this class, emergent behaviours also appear and some of them also seem to be unavoidable (though no proof was found).

In this note we present the simplest known generalised ant that has at least two distinct asymptotic behaviours.
Indeed, contrary to what (non-extensive) simulations may suggest, and contrary to a somewhat common belief, the $LLLR$ ant has at least two distinct asymptotic behaviours.
In Section \ref{sec:def} we define the $LLLR$ ant, in Section \ref{sec:highways} we present the two known highways for the $LLLR$ ant, and in Section \ref{sec:further} we have a discussion on the related results and further work.

The behaviours we present can be viewed on our online simulator at \href{https://lutfalla.fr/ant}{lutfalla.fr/ant} whose source code is publicly available \cite{lutfalla2025}.

\section{The $LLLR$ ant}
\label{sec:def}
\begin{figure}[htp]
  \includegraphics[width=\textwidth]{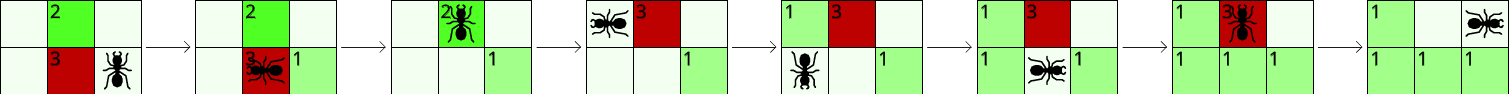}
  \caption{A simple example with $7$ steps of the dynamics of the $LLLR$ ant.}
  \label{fig:def}
\end{figure}

The $LLLR$ ant is a simple automaton evolving with a grid configuration $c\in \{0,1,2,3\}^{\mathbb{Z}^2}$.
When the ant arrives on a cell in state $0$, $1$ or $2$ it turns counterclockwise, increases the cell state by $1$ modulo $4$ and moves one step forward.
When the ant arrives on a cell in state $3$ it turns clockwise, increases the cell state by $1$ modulo $4$ and moves one step forward, see Fig. \ref{fig:def}.

We call \emph{ant configuration} a triplet $(c, (i,j), q)$ where $c\in \{0,1,2,3\}^{\mathbb{Z}^2}$ is the grid configuration, $(i,j)\in \mathbb{Z}^2$ is the ant position, and $q\in\{\Nant,\Eant,\Sant,\Want\}$ is the ant direction.

The ant $LLLR$ is formally a transition function $T_{LLLR}$ over ant configurations defined as 
 $T_{LLLR}(c,(i,j),q)=(c',(i',j'),q')$ with :
\begin{itemize}
\item denoting $k=c_{i,j}$, if $k=3$ then $q':= \turnright(q)$, otherwise $q':=\turnleft(q)$
\item $(i',j'):=(i,j)+q'$
\item $c'_{i,j}:= c_{i,j} + 1 \mod 4$
\item $\forall (k,l)\neq(i,j), c'_{k,l}:=c_{k,l}$
\end{itemize}

We call \emph{trace} of ant $LLLR$ from configuration $C=(c,(i,j),q)$ the sequence of symbols (grid cell states) that the ant encounters when evolving from $C$, and \emph{trajectory} of $LLLR$ from $C$ the sequence of coordinates of the ant when evolving from $C$.
The trace, which is an infinite word on alphabet $\{0,1,2,3\}$, contains all the information about the behaviour of the ant and the initial configuration (or at least on the portion of $\mathbb{Z}^2$ that was visited by the ant).

We say that a grid configuration $c\in\{0,1,2,3\}^{\mathbb{Z}^2}$ is \emph{finite} if there are only finitely many non-zero cells in $c$.

\begin{definition}[Highway behaviour]
We say that ant $LLLR$ is in a \emph{highway} of period $n$ and drift $(a,b)$ from configuration $C=(c,(i,j),q)$ if :
\begin{itemize}
\item $c$ is finite, that is : $|\{(i,j)\in \mathbb{Z}^2, c_{i,j} \neq 0\}|< +\infty$
\item the trace $x\in \{0, 1, 2, 3\}^\mathbb{N}$ of the ant from $C$ is $n$-periodic\footnote{We also assume that the trace $x$ is not $k$-periodic for $k<n$.}, that is : for any $i$ we have $x_{i+n} = x_i$
\item the trajectory $\mathbf{y} \in (\mathbb{Z}^2)^\mathbb{N}$ of the ant from $C$ is $n$-periodic modulo an $(a,b)$-drift, that is : for any $i$ we have $\mathbf{y}_{i+n} = \mathbf{y}_i + (a,b)$
\end{itemize}
\end{definition}

Remark that the ant dynamics commutes with rotations of the whole ant configuration (grid configuration, ant position and ant direction) by angles multiple of $\tfrac{\pi}{2}$.
Hence if the configuration $C$ starts a highway of period $n$ and drift $(a,b)$, then its image $C'$ by a rotation of angle $\tfrac{\pi}{2}$ starts a highway of period $n$ and drift $(-b,a)$.
In the usual case of diagonal highways, where we have $|a|=|b|$, this means that the drift can be defined up to $\pm$ on both coefficients.

By analogy, we say that the ant starts a highway of period $n$ and drift $(a,b)$ from a finite pattern $P$ if it start such a highway from the finite configuration obtained by perturbating the $0$-uniform configuration by pattern $P$ at the origin.

We consider that two highways for ant $LLLR$ are equivalent when their trace and trajectory (up to a global $\tfrac{k\pi}{2}$ rotation) are identical up to removing a finite prefix from both.
And we consider highways up to this equivalence relation.

We also say that the ant eventually enters or reaches a highway from configuration $C$ if there exists an integer $N$ such that the ant starts a highway at configuration $T_{LLLR}^N(C)$.

\section{Asymptotic behaviours of the $LLLR$ ant}
\label{sec:highways}
From initially finite configurations, the ant $LLLR$ reaches at least two distinct highways:
\begin{itemize}
\item a simple highway of period $52$ and drift $(\pm 2,\pm 2)$ (which resembles known highways for the $L^{5}R$ ant and all $L^{2k+1}R$ ants with $k>1$),
\item and a longer and more complex highway of period $156$ and drift $(\pm 2, \pm 2)$ (which is quite unique).
  \end{itemize}
In our experiments, when starting from a random pattern of size $11\times 11$ drawn uniformly at random, the ant reaches the simple highway in $99.88\%$ of cases and the complex highway in $0.12\%$ of cases.

As explained above, the trace of the ant contains all of the information about the dynamics of the ant and the grid configuration, or at least the part of the configuration that is visited and up to a global $\tfrac{k\pi}{2}$ rotation of the trajectory.
Therefore, we describe these two highways through the periodic word in the trace and a minimal seed pattern that starts the highway.

\subsection{The simple highway of period 52}
\begin{figure}[h]
  \includegraphics[width=\textwidth]{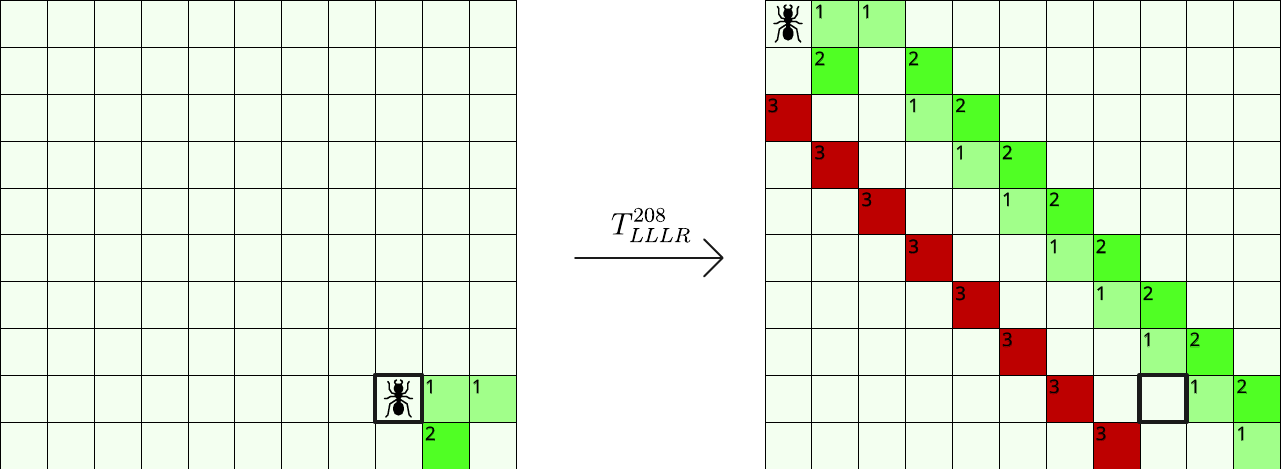}
  \caption{The pattern $P_{52}$ starting the simple highway of period $52$ and drift $(-2,2)$ for the $LLLR$ ant (left), and the configuration reached from $P_{52}$ after $208$ steps, \emph{i.e.}, four periods (right).\\
    The cell marked with bold boundary is the origin cell.
  }
  \label{fig:p52}
\end{figure}

From pattern $P_{52}$ of Fig.~\ref{fig:p52} the $LLLR$ ant starts a highway of period $52$ and drift $(\pm 2, \pm 2)$ ilustrated at \href{https://lutfalla.fr/ant/highway.html?antword=L2K1R}{lutfalla.fr/ant/highway.html?antword=L2K1R}, the periodic word of the trace is $u_{52}$ with
\[ u_{52} := \texttt{0000111122223100021113201033230000111122223200033313}\]
The speed of this simple highway is $v_{52}:=\frac{2\sqrt{2}}{52} = \frac{\sqrt{2}}{26}$.

\begin{figure}[h]
  \includegraphics[width=\textwidth]{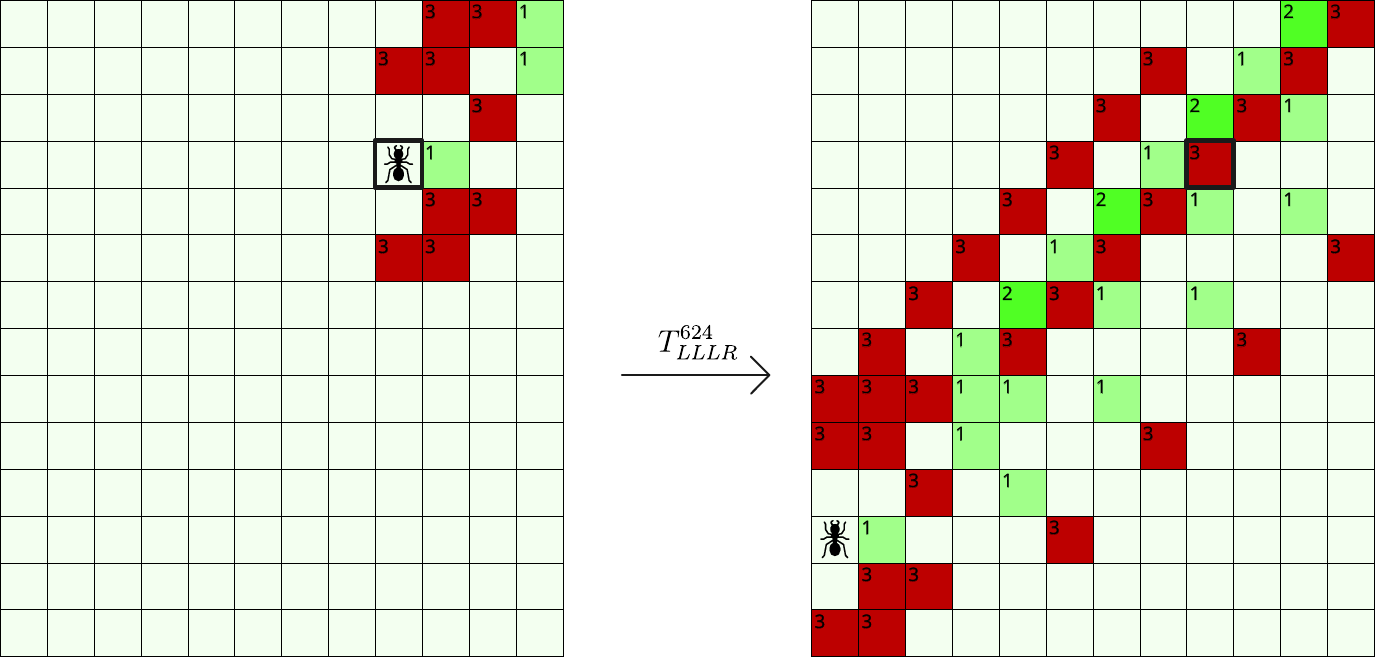}
  \caption{The pattern $P_{156}$ starting the complex highway of period $156$ and drift $(-2,-2)$ for the $LLLR$ ant (left), and the configuration reached from $P_{156}$ after $624$ steps, \emph{i.e.}, four periods (right).\\
    The cell marked with bold boundary is the origin cell.
  }
  \label{fig:p156}
\end{figure}

\subsection{The complex highway of period 156}
From pattern $P_{156}$ of Fig.~\ref{fig:p156} the $LLLR$ ant starts a highway of period $156$ and drift $(\pm 2,\pm 2)$ illustrated at \href{https://lutfalla.fr/ant/highway.html?antword=L3R156}{lutfalla.fr/ant/highway.html?antword=L3R156}, the periodic word of the trace is $u_{156}$ with
\begin{align*} u_{156} := & \texttt{000011112222310002111333230000111122223233230000111122223003}\\
  & \texttt{000011112222331030000111122223313133033312202303}\\
  & \texttt{000011112222303301101221233030323030000111122223}
\end{align*}
The speed of this complex highway is $v_{156}:=\frac{2\sqrt{2}}{156}= \frac{\sqrt{2}}{78} = \frac{v_{52}}{3}$.

\subsection{Experiments and statistics}
As explained above, these two behaviours are relatively well known (and might be considered folklore) and were found by experimentation.

We ran some intensive computations on several generalised ants, some of them being discussed in \cite{gajardo2025}. In those simulations, we choose (pseudo-)randomly a finite initial configuration by perturbating the $0$-uniform configuration by a $11\times 11$ pattern centred on the origin drawn uniformly at (pseudo-)random.
We then run the $LLLR$ ant for $10^{5}$ time steps and analyse the trace to determine if it has reached a highway.

We ran $288358500$ experiments for the $LLLR$ ant, of which $100\%$ had reached a highway after $10^5$ steps. Of those $99.88\%$ reached the simple highway of period $52$ and only $0.12\%$ reached the complex highway of period $156$.

\begin{figure}[h]
  \includegraphics[width=\textwidth]{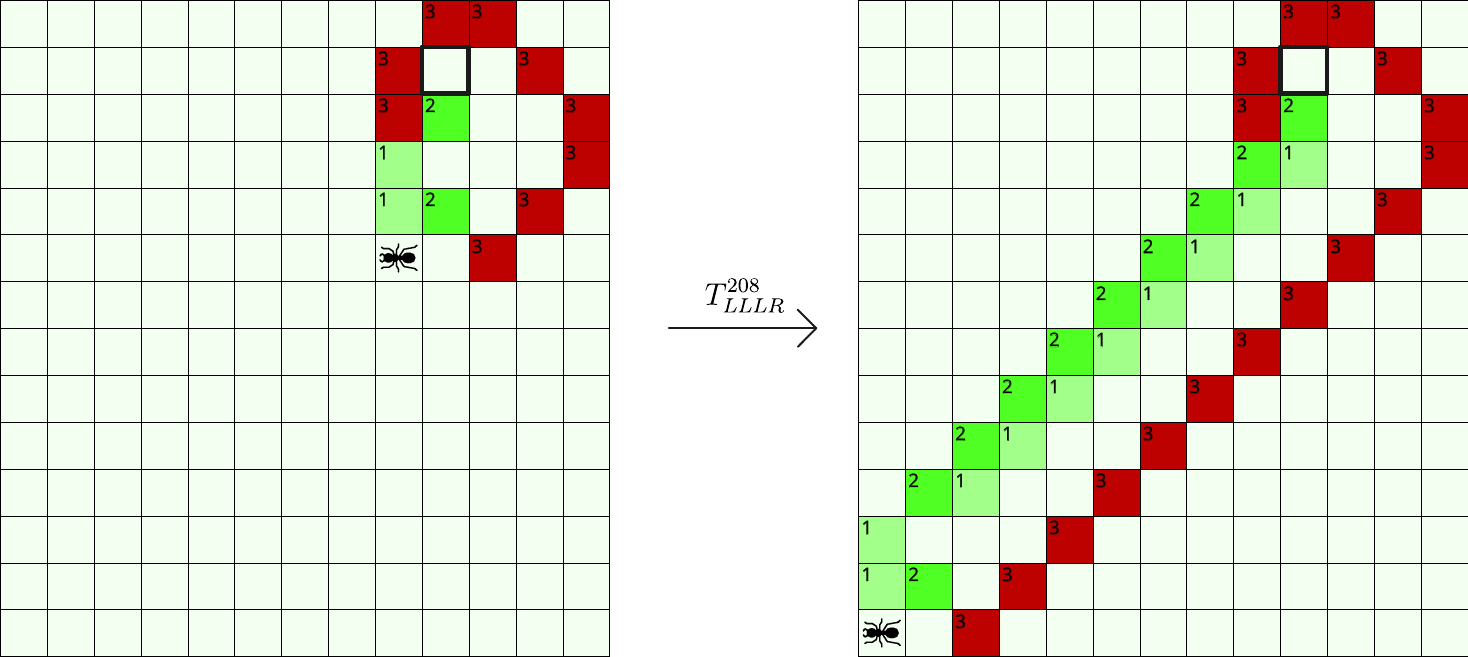}
  \caption{The configuration reached from the $0$-uniform configuration after $105$ steps (left), and then $208$ steps latter (right), that is, after four periods of the simple highway which was reached.\\
    The cell marked with bold boundary is the origin cell.
  }
  \label{fig:from0uniform}
\end{figure}

Note also that from the $0$-uniform configuration, the $LLLR$ ant reaches the simple highway after $105$ steps, see Fig. \ref{fig:from0uniform} or \href{https://lutfalla.fr/ant/gl_ant.html?antword=LLLR}{lutfalla.fr/ant/gl\_ant.html?antword=LLLR}.

\section{Related results and further work}
\label{sec:further}
For any $k>0$, the $L^{2k+1}R$ ant has a highway of period $32k+20$ and drift $(\pm 2, \pm 2)$.
More precisely, the ant $L^{2k+1}R$ has $k$ distinct highways of same period $32k+20$ and drift $(\pm 2, \pm 2)$ but with different traces, see \cite[Remark 3]{gajardo2025}.
The simple highway of period $52$ for the $LLLR$ ant is simply the case $k=1$ of this behaviour.

On the other hand, the complex highway of period $156$ for the $LLLR$ ant is quite unique as no similar behaviour was discovered for $L^{2k+1}R$ ants with $k>1$ despite billions of experiments and some precise study and decomposition of this complex highway.

As this complex highway is already quite rare with only $0.12\%$ of our experiments reaching it, it might be possible that a similar highway exists for the $L^{5}R$ ant (and other $L^{2k+1}R$) but with frequencies so negligible that we never encountered them. We are therefore actively continuing our research for such complex highways for the $L^{5}R$ ant.

Note also that it was, at some point, believed that all $L^+R$ ants have a unique highway behaviour from initially finite configurations (up to a permissive equivalence relation on highways, considering highways of same speed and similar trajectories as equivalent). The $LLLR$ ant is a counter-example to this belief.

As a final note, remark that the quiescent state (or background state) is very important.
Indeed the asymptotic behaviour from finite configurations that we observe for the $LLLR$ ant and for the $LLRL$ ant are very dissimilar, though a finite configuration for $LLRL$ can be seen as a $1$-finite configuration (a finite number of non-$1$ cells) for $LLLR$.
For the $LLRL$ ant we observe an overwhelmingly dominant highway of period $384$ and four extremely rare highways of periods $308$, $380$, $388$ and $928$ (for more details see \cite[\S 3]{gajardo2025}).
Those highways bear no resemblance to the highways we observe on the $LLLR$ ant.

\printbibliography
\end{document}